\documentclass[fleqn,usenatbib]{mnras}

\usepackage{newtxtext,newtxmath}
\usepackage{color}

\usepackage[T1]{fontenc}
\usepackage{ae,aecompl}
\usepackage{multirow}

\newcommand{\rev}[1]{\textcolor{black}{#1}}

\definecolor{orange}{rgb}{1,0.5,0}

\newcommand{\msunyr}{$M_\odot$ yr$^{-1}$}
\def\msun{\rm{\,M_{\odot}}}

\def\msunh{\,h^{-1}\rm{\,M_{\odot}}}

\def\kms{\rm{\,km\,s^{-1}}}

\usepackage{graphicx}	
\usepackage{amsmath}	
\usepackage{amssymb}	

\title[BCG Evolution]{BCG Mass Evolution in Cosmological Hydro-Simulations}

\author[Ragone-Figueroa et al.]{
C.\ Ragone-Figueroa$^{1,2},$\thanks{E-mail: cin@oac.unc.edu.ar}
G.L.\ Granato$^{2,1}$,
M.E.\ Ferraro$^{3}$,
G. Murante$^{2,1}$,
V.\ Biffi$^{2,4}$,\\~\\
{\rm {\LARGE
S.\ Borgani$^{4,2}$, S. Planelles$^5$ and E.\ Rasia$^{2}$}}
\\
$^{1}$ Instituto de Astronom\'ia Te\'orica y Experimental (IATE), Consejo Nacional de Investigaciones Cient\'ificas
y T\'ecnicas de la Rep\'ublica \\
Argentina (CONICET), Observatorio Astron\'omico, Universidad Nacional de C\'ordoba, Laprida 854, X5000BGR C\'ordoba, Argentina\\
$^{2}$ INAF, Osservatorio Astronomico di Trieste, via Tiepolo 11, I-34131, Trieste, Italy \\
$^3$ FaMAF, Facultad de Matem\'atica, Astronom\'ia, F\'isica y Computaci\'on, Av. Medina Allende s/n , Ciudad Universitaria,\\ CP:X5000HUA C\'ordoba, Argentina. \\
$^4$ Dipartimento di Fisica dell' Universit\`a di Trieste, Sezione di Astronomia, via Tiepolo 11, I-34131 Trieste, Italy \\
$^5$ Departament d'Astronomia i Astrof{\'i}sica, Universitat de Val\`encia, c/ Dr. Moliner, 50, 46100 - Burjassot (Valencia), Spain\\
}

\date{Accepted 2018 June 17. Received 2018 June 6; in original form 2018 March 15.}

\pubyear{2018}

\begin{document}
\label{firstpage}
\pagerange{\pageref{firstpage}--\pageref{lastpage}}
\maketitle

\begin{abstract}

We analyze the stellar growth of Brightest Cluster Galaxies (BCGs) produced by cosmological zoom-in hydrodynamical simulations of the formation of {\it massive} galaxy clusters.
The evolution of the stellar mass content is studied considering different apertures, and  tracking backwards either the main progenitor of the $z=0$ BCG or that of the cluster hosting the BCG at $z=0$. Both methods lead to similar results up to $z \simeq 1.5$. The simulated BCGs masses at $z=0$ are in  agreement with recent observations.
In the redshift interval from $z=1$ to $z=0$  we find growth factors 1.3, 1.6 and 3.6 for stellar masses within 30kpc, 50kpc and 10\% of $R_{500}$ respectively. The first two factors, and in general the mass evolution in this redshift range, are in agreement with most recent observations.
The last larger factor is similar to the growth factor obtained by a semi-analytical model (SAM).
Half of the star particles that end up in the inner 50 kpc was typically formed by redshift $\sim$ 3.7, while the assembly of half of the BCGs stellar mass occurs on average at lower redshifts $\sim 1.5$. This assembly redshift correlates with the mass attained by the cluster at high $z \gtrsim 1.3$, due to the  broader range of the progenitor clusters at high-$z$.
The assembly redshift of BCGs decreases with increasing apertures. Our results are compatible with the {\it inside-out} scenario. Simulated BCGs could lack intense enough star formation (SF) at high redshift, while possibly exhibit an excess of residual SF at low redshift.

\end{abstract}




\section{Introduction}
\label{sec:intro}
A brightest cluster galaxy (BCG) is defined as the most optically luminous galaxy in a cluster. BCGs are found at or near the cluster centre, and they are characterized by several properties that distinguish them from other bright galaxies \citep[e.g.][]{vonderlinden2007,bernardi2009}. At the most basic level, it was recognized more than 40 years ago that the BCG population is inconsistent with the global luminosity function of galaxies \citep[see][and references therein]{tremaine1977}.
It seems therefore natural to consider peculiar formation paths for these impressive objects.

In particular, two processes specifically related to the BCGs central position, could in principle contribute  significantly to the mass growth at relatively late times, once their host halos are virialized. These are {\it cooling flows} \citep{cowie1977,fabian1977} and {\it galactic cannibalism} \citep{ostriker1975,white1976}.
Cooling flows refer to the sinking of hot intra cluster medium (ICM) toward the center of the potential well, due to radiative cooling and consequent decrease of  pressure support. Unless some heating mechanism completely counterbalances the radiative losses, cooling flows could drive to the BCG region star forming gas at important rates, well exceeding  hundreds of \msunyr.
However, starting from the early 2000s, spatially-resolved X--ray spectroscopic studies with XMM-Newton and Chandra showed that local galaxy clusters lack signatures of the huge cooling flows predicted by classical models \citep[e.g.][]{peterson2003}. This {\it cooling flow problem} has been one of the main motivations to include a treatment of super massive black hole (SMBH) growth and of the associated feedback in models of galaxy formation, both semi-analytic \citep[e.g.][]{granato2004,croton2006,bower2006} and based on simulations \citep[e.g.][]{sijacki2007}.
{\it Galactic cannibalism} is instead the merging with satellite galaxies. It is caused by dynamical friction which gradually sinks galaxies toward the cluster central region. Since the process timescale is inversely proportional to the satellite mass, it promotes preferential accretion of massive satellites. Most recent models ascribe to this effect the bulk of the BCGs mass growth after $z\sim$ 1. According to these models, the mergers are expected to be {\it dry}, i.e.\ gas poor and involving negligible star formation. However, the predicted growth seems to be somewhat larger than observed. This possible tension could be worsened by recent claims, based on FIR data, that BCGs exhibit a star formation activity at $z \lesssim 1$ more significant than previously thought, and thus contributing in a non-negligible way to their late mass growth \cite[][]{bonaventura2017}.

During the last decade, the increasing availability of relatively large samples of galaxy clusters up to $z=1$-$1.5$ has triggered  plenty of studies on the ``late" mass growth of BCGs. However the published results are quite contradictory.
Some of these works claim for little or no change since  z $\sim$ 1 \citep[][see also \citealt{collins1998}]{whiley2008,zhang2016} or since z $\sim$ 1.5 \citep{collins2009,stott2010}, while others find a growth factor of $\sim$ 2 between $z=1$ and $z=0$ and suggest a stall at z$\lesssim 0.5$ \citep{lidman2012,lin2013, bellstedt2016,lin2017}.
However, this lack of consensus could arrive at least in part from the different methods that are used to estimate the mass of the BCGs and also from dissimilarities in the sample selection. \rev{Most recent work apply some selection criteria at different redshift in such a way that samples at different redshift mimic as much as possible an evolutionary sequence.}

\rev{A different approach used to estimate the fraction of BCG growth {\it due to mergers} is based on counting the number of close galaxies that can be expected to merge with them before the present epoch. A few assumptions are required to perform these estimates. Projection effects need to be accounted for as well as the fraction of stars that are stripped to the Intra Cluster Light (ICL) component during the mergers.  Also the merger timescale is uncertain \citep{liu2009}. Results of simulations are usually employed to get constraints on these points. The first two of them, when  not taken into account, can in principle lead to overestimate the growth. On the other hand, these estimates do not include the mass growth due to star formation, in some cases consider only major mergers \citep[e.g.][]{liu2009,mcintosh2008}, and in others do not evaluate the contribution of galaxies which are not close to the BCG at the epoch of observation \citep[e.g.][]{edwards2012,burke2015}. For these latter reasons they should be considered as lower limit to the total growth.}

From a theoretical point of view, most studies on the mass evolution of BCGs have been performed by means of semi-analytic models (SAM) of galaxy formation. The general trend has been in the sense of a downward revision of the mass growth between $z=1$ and $z=0$, likely driven by the data suggesting less evolution than early predictions. The first specific estimates were presented by \cite{aragon1998}, using both the Durham as well as the Munich SAM at the time \citep[][respectively]{baugh1996,kauffmann1998}. When adopting the then standard $\Omega_m=1$ CDM cosmological model, the
predicted growth factor was in both cases $\sim 4$. However in that cosmological model the development of cosmic structures is delayed with respect to the $\Lambda$CDM, which became standard a few years later.

However, \cite{delucia2007} still found an average growth of a factor $\sim 3$ for BGCs in their much later rendition of the Munich SAM. This model was run on merger trees extracted from the Millenium $\Lambda$CDM simulation \citep{springeletal2005} and included also cooling flow suppression due to AGN feedback, in order to limit their late growth.
However, analyzing a quite similar version of the Munich SAM, \cite{tonini2012} found that BCGs increase their masses by a factor of 2-3 since $z\sim1.6$, and that most of this growth occurs at $z \gtrsim 1$.

A possible limitation in studies of BCGs formation with SAMs could be that in this theoretical technique their peculiarity is imposed by construction. Indeed the fact that shock heated gas cools radiatively and condenses only onto central galaxies is one of the assumptions of these models.

Moreover, usually only mergers between central galaxies and satellites are taken into account, neglecting satellite-satellite mergers. Furthermore, SAMs lack detailed spatial information on the model galaxy, apart from some analytical estimate of typical scale-length and an {\it assumed} density profile. Therefore their true prediction is the ``total" masses of galaxies. This is what usually has been compared with observations, which instead estimate masses (from luminosities) within a given aperture (see discussion in \citealt{whiley2008} and also Section \ref{sec:evomass}). A somewhat related problem is that only very recently a treatment of the formation and evolution of the intracluster light (ICL) begun to be included in SAMs. This slows down the late $z\lesssim 1$ BCGs mass growth, by decreasing progressively the masses of accreting satellites (\citealt[][]{contini2014}; see also \citealt[][]{monaco2006}).

A different modeling method to study the late mass growth of BCGs has been employed by
\cite{dubinski1998}, albeit  adopting a closed CDM cosmological model, and more recently by
\cite{ruszkowski2009} and \cite{laporte2015}. In this approach, gravity only simulations are employed. Below a certain redshift $z \sim 2$, the dark matter subhalos are populated by stellar systems with properties matching as much as possible real galaxies, and the simulation is continued down to $z=0$. Another scheme, also based on pure N--body simulations, has been employed by \cite{laporte2013}. In this case at $z=2$ the simulation particles are weighted to represent realistic stellar density profiles for the initial galaxies, and the subsequent evolution of ``galaxies" is analyzed adopting these weights.
In both cases, the underlying assumption is that the assembly of the inner regions
of galaxy clusters is only driven by dry mergers. Some observations \citep[e.g.][]{cooke2016} could support such hypothesis at $z \lesssim 1$. However several other indications point to a different view, in particular at $z \gtrsim 0.5$, wherein the gas component, the associated star formation and SMBH activity still have a significant role \citep{tran2010,santos2015,mcdonald2015,bonaventura2017}.

In conclusion, a comprehensive theoretical description of BCGs formation calls for hydrodynamical simulations in a cosmological context, such as those we use in the present work.
To the best of our knowledge, there are no specific studies based on cosmological hydrodynamical simulations devoted to investigate the mass growth of BCGs. However \cite{martizzi2016}, in a more general study dedicated to the evolution of the baryonic component
in AMR simulations of 10 galaxy clusters with $M_{vir} \simeq 10^{14} \mbox{M}_\odot$, reported a moderate mass growth $\lesssim 2$ within 50 kpc since $z=1$.

The paper is organized as follows: in Section \ref{sec:method} we give a brief description of the simulated clusters and of the analysis methods. In Section \ref{sec:results} we present and discuss our results which are then summarized in Section \ref{sec:summary}.

\section{Method}
\label{sec:method}
\subsection{The Simulated Clusters}
\label{sec:thesimulatedclusters}

\begin{figure*}
\includegraphics[width=8.5cm, height=8cm]{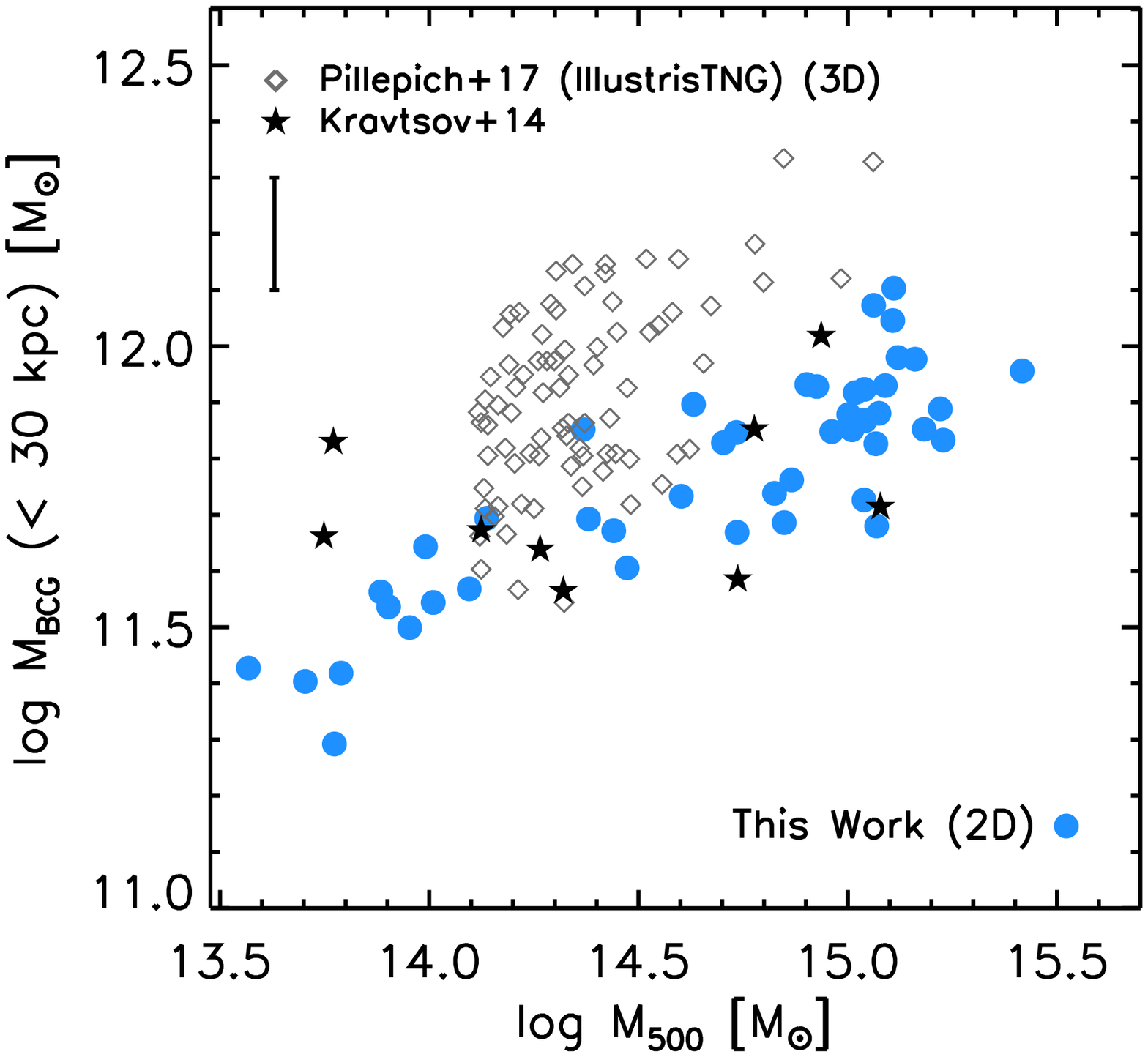}
\includegraphics[width=8.5cm, height=8cm]{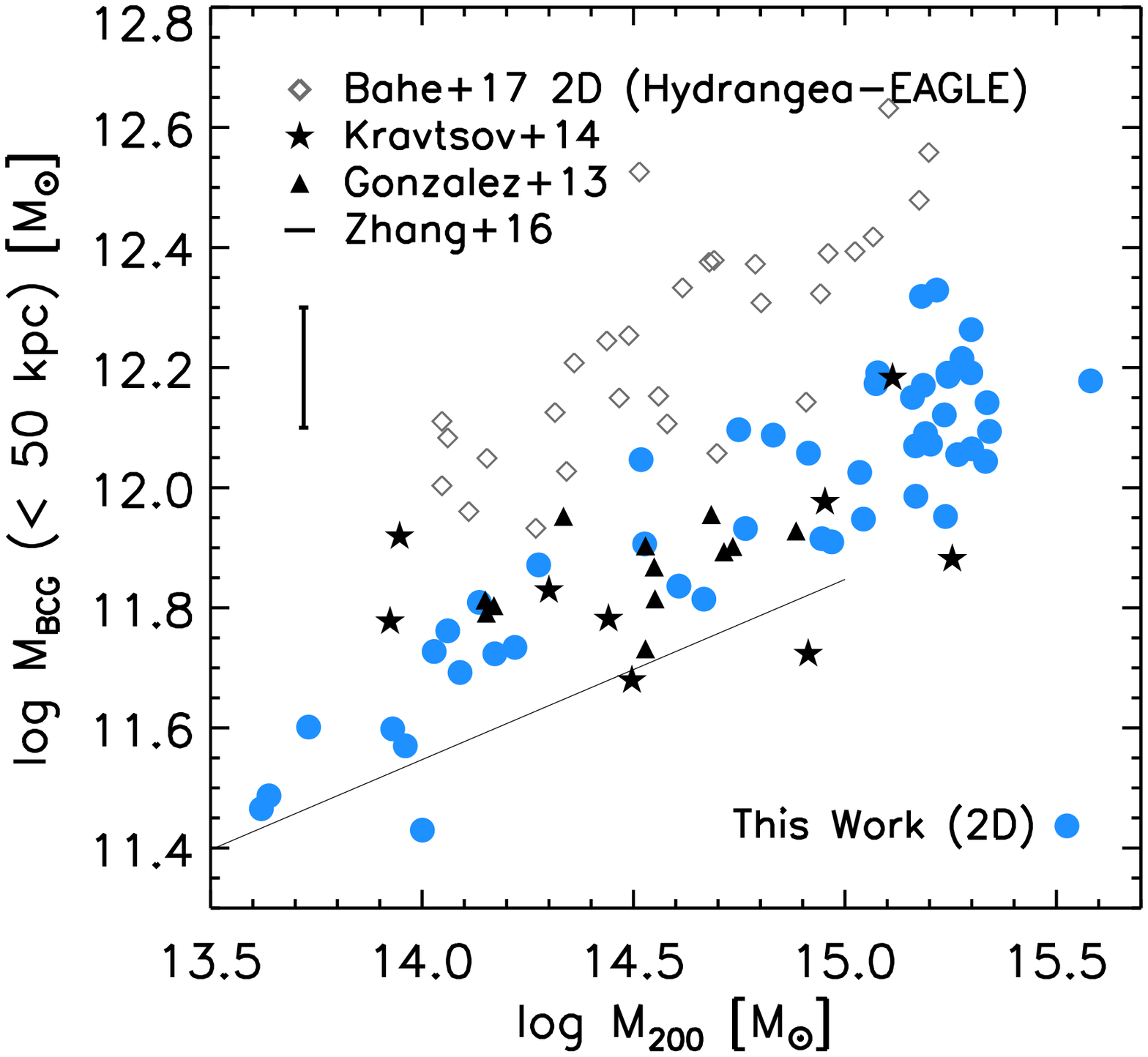}
\caption{Cluster mass - BCG mass relation obtained in this work (blue filled circles) compared to other simulated results (empty diamonds) and observations (filled stars, triangles and solid line).
\rev{The vertical bars show the maximum overall error associated to the \citet{kravtsov2014} data.}
In the left (right) panel BCG masses are computed within 30 (50) projected kpc while the cluster mass is $M_{500}$ ($M_{200}$). Only in this figure we use the extended sample of 48 clusters (see text).
 } \label{fig:mclus_mbcg_fig}
\end{figure*}

We re-run  and analyze zoom-in simulations for the formation of massive galaxy clusters, from a set of initial conditions already presented in previous papers. We describe briefly in this section  the main aspects of these simulations, focusing on the differences with respect to previous runs. We refer the reader mostly to \cite{ragone2013}
for further numerical details.

We simulate 24 zoomed-in Lagrangian regions with a custom version of the {\footnotesize GADGET-3} code~\cite[][]{springel2005}. These regions have been selected from a parent gravity only simulation of a 1 $h^{-1}$Gpc box, and surround all the 24 dark matter halos with final $M_{200}>8\times 10^{14}\, \msunh$ \footnote{$M_{200}$ ($M_{500}$) is the mass enclosed by a sphere whose mean density is 200 (500) times the critical density at the considered redshift.}. Each region was re-simulated at higher resolution including baryonic physics.
The adopted cosmological parameters are: $\Omega_{\rm{m}} = 0.24$, $\Omega_{\rm{b}} = 0.04$, $n_{\rm{s}}=0.96$, $\sigma_8 =0.8$ and $H_0=72\,\kms$\,Mpc$^{-1}$.

The mass resolution for the DM and gas particles is $m_{\rm{DM}} = 8.47\times10^8 \, \msunh$ and $m_{\rm{gas}} =1.53\times10^8\, \msunh$,
respectively. For the gravitational force, a Plummer-equivalent softening length of $\epsilon = 5.6\, h^{-1}$\,kpc is used for DM and gas particles, whereas $\epsilon = 3\, h^{-1}$\,kpc for black hole and star particles. The DM softening length is kept fixed in comoving coordinates for $z>2$ and in physical coordinates at lower redshift.

We used the SPH formulation by \cite{beck2016}, that includes artificial thermal diffusion and a higher-order interpolation kernel, which improves the standard SPH performance in its capability of treating discontinuities and following the development of gas-dynamical instabilities.

Our set of simulations include a treatment of all the unresolved baryonic processes usually taken into account in galaxy formation simulations. For details on the adopted implementation of cooling, star formation (SF), and associated feedback, we refer the reader to \cite{ragone2013}. In brief, the model of SF is an updated version of the implementation by~\cite{springel2003}, in which gas particles with a density above $0.1\,$cm$^{-3}$ and a temperature below $2.5\times 10^5$\,K are classified as multiphase. Multiphase particles comprise a cold and a hot-phase, in pressure equilibrium. The cold phase is the star formation reservoir. Metallicity dependent cooling is implemented following the approach by \cite{wiersma2009}. The production of metals is followed according to the model of stellar evolution originally implemented by \cite{tornatore2007}.

A full account of the AGN feedback model can be found in Appendix A of \cite{ragone2013}. The main difference introduced here with respect to that work is the distinction between a {\it cold mode} and a {\it hot mode} gas accretion. This was inspired by the result of high resolution AMR simulations of the gas flowing to SMBHs \citep{gaspari2013}. In practice, we apply the formula for the $\alpha$-modified Bondi accretion rate to the hot and cold gas components separately;
\begin{equation}
  \dot{M}_{Bondi,\alpha} = \alpha \,   \frac{4 \pi G^2M^2_{BH} \rho}{\left(c^2_s+v^2_{BH}\right)^{2/3}}
\label{eq:bondi}
\end{equation}
where $M_{BH}$ is the BH mass, $v_{BH}$ its velocity relative to the surrounding gas bulk motion, $c_s$ is the sound velocity of the gas sorrounding the BH and $\rho$ its density.
The threshold between the two gas components is set at $T=5\times 10^5$ K, and the adopted values of the fudge factor $\alpha$ are 100 and 10 respectively, as suggested by the results of \cite{gaspari2013}. We also introduced a small modification which substantially improves the association of SMBH particles with the stellar system in which they were first seeded, with respect to all previous versions of our simulations based on the same set of initial conditions. As detailed in \cite{ragone2013}, this is mostly attained by moving at each time-step the SMBH at the position of the most bound particle within its softening length. However now the search of the minimum is done taking into account only those particles whose  velocity differs less than a given limiting value from the BH velocity. This is to avoid jumps of the SMBH from its original stellar system to another one during a close flyby. After carrying out a number of numerical experiments, we verified that threshold relative velocities between 100 and 200 km/s do the best job, at least at the present resolution. These values are all but surprising, since they are of the same order of the velocity dispersion of stars in galaxies, but much smaller than the velocity dispersion of galaxies in a cluster.
\rev{A fraction $\epsilon_r$ of the energy associated to the gas mass accreting onto the SMBH is assumed to be radiated away, and a fraction $\epsilon_f$ of this radiation is then thermally coupled to the surrounding gas. These parameters are calibrated to reproduce the observed scaling relations of SMBH mass in spheroids. Here we set $\epsilon_r=0.07$ and $\epsilon_f=0.1$. We also assume a transition from a {\it quasar mode} to a {\it radio mode} AGN feedback when the accretion rate becomes smaller than a given limit, $\dot{M}_{BH}/\dot{M}_{Edd} = 10^{-2}$. In radio mode we increase the feedback efficiency $\epsilon_f$ to 0.7.}

\subsection{Cluster selection and BCGs identification}
\label{sec:clussel}
The identification of host clusters has been done by running a FoF algorithm in the high resolution regions, which links DM particles using a linking length of 0.16 times the mean inter-particle separation.
The main body of this work is done using the 24 central clusters
of the resimulated regions mentioned above. These massive clusters have masses at present time $ 0.7 \lesssim M_{500}/10^{15}\msun \lesssim 2.6 $, with a median of $\sim 1.2 \times 10^{15} \mbox{M}_\odot$.

Only in Section \ref{subsec:mbcgmclus} do we expand the cluster sample to 48 by selecting the two most massive clusters in each Lagrangian region. In this case the mass range is $ 0.03 \lesssim M_{500} / 10^{15} \msun \lesssim 2.6 $ and the median $\sim 7 \times 10^{14} \msun$.

The center of BCGs at $z=0$ is associated with the position of the particle with minimum potential in the main subhalo of the cluster, as identified by Subfind (\citealt{dolag2009}). This main subhalo contains, besides the  star particles that compose the BCG, all particles that were not associated to any other subhalo.

We consider two possibilities to follow back in time the BCG mass assembly. In the first one we search for its main progenitor from one output snapshot to the previous one, along our sequence of snapshots equally spaced in expansion factor, with $\delta a \simeq 10^{-2}$. However it may happen that the main progenitor of the $z=0$ BCG is not a BCG at higher redshift, with increasing probability for this to happen at higher redshift.
This possibility has to be evaluated when confronting with observational estimates of BCG growth factors. Indeed the latter are performed by comparing masses of BCGs selected at different redshifts, from clusters that approximate as well as possible an evolutionary sequence. In this case a more proper comparison with the simulations can be done by considering the assembly history of the BCG in the main progenitor of the cluster. In practice, for our sample of simulations the two choices give on average almost indistinguishable results up to $z\sim 2$, which safely exceeds the redshift for which data are available (see Section \ref{sec:evomass}).

\subsection{Definitions of BCG masses}
In this work we compute BCG masses within spherical radii of 30kpc, 50kpc and 0.1$R_{500}$. From now on, we refer to these masses as M30, M50 and M10\% respectively. The purpose of using different mass definitions is to study the dependence of results on the way mass is computed.

Masses measured within fixed physical radii has been often advocated and employed, in order to cleanly compare  models with observations \citep[e.g.][]{kravtsov2014}. This point is particularly delicate for BCGs, due to the ill defined transition to the ICL. It is worth pointing out that according to \cite{stott2010}, BCG luminosities obtained within a 50 kpc aperture differ from those computed with {\tt MAG\_AUTO} by less than 5\%. {\tt MAG\_AUTO} uses a kron-like aperture and is often present in observational analyses.

On the other hand, in analyzing simulations a practical, albeit not well justified, choice has been often to define as {\it galaxy} the baryonic component within a certain fraction of the virial radius \citep[e.g.][]{scannapieco2012}. On the observational side \cite{kravtsov2014} evaluate masses of BCGs within 10\% of $R_{500}$. It is therefore interesting to consider also the mass contained within this radius. In our sample of massive clusters, $0.1 R_{500}$ amounts on average to 160, 60 and 20 physical kpc at redshift 0, 1 and 2 respectively. Therefore it is the largest aperture we consider typically up to redshift $\sim 1.2$. We anticipate that we find a striking similarity between the evolution of M10\% in our simulations and that predicted for the total mass by the \cite{delucia2007} SAM (Section \ref{sec:evomass}).

\section{Results}
\label{sec:results}

\subsection{BCG mass vs. Halo Mass at z=0}
\label{subsec:mbcgmclus}
Considering that galaxies evolve within their host dark matter halos, it is expected a connection between the stellar mass of central galaxies and the host halo mass.
This correlation has been studied in both observational (e.g. \citealt{stott2012}, \citealt{kravtsov2014}, \citealt{bellstedt2016}) and theoretical works (e.g. \citealt{stott2012}, \citealt{pillepich2017}, \citealt{bahe2017}).

In Fig. \ref{fig:mclus_mbcg_fig} we use our extended sample of 48 clusters to compare the relationship between simulated BCG and halo masses at redshift zero with results from both observations and other numerical works. The left and right panels show masses measured within 30 and 50 projected kpc respectively. We find, for the same halo mass, BCG stellar masses smaller by a factor 2 to 3 than those obtained by other state of the art hydrodynamical simulations, such as IllustrisTNG \citep{pillepich2017} and EAGLE \citep{bahe2017}. As a consequence of the smaller masses, our BCGs turn out to be in better agreement with the observational estimates \citep{zhang2016, kravtsov2014, gonzalez2013}.
\rev{The smaller BCG masses in our simulations could be due to our coarser resolution. However we point out that we run a few test cases with mass resolution improved by a factor 3, obtaining relatively stable BCG stellar masses within $\sim 10\%$ and, more importantly, lacking a systematic trend for a mass increase or decrease as resolution is improved.  Moreover, several other recent  studies, most of which are characterized by a resolution similar to our own,
over-predict anyway the present-day BCG stellar masses, even including AGN feedback \citep[see discussion in][]{ragone2013}. \cite{bahe2017} as well as \cite{ragone2013} suggested that this long-standing problem could be resolved by more realistic AGN feedback models, which should be
more efficient at expelling gas from massive halos at high redshift. In our case, most of the improvement with respect to \cite{ragone2013} is due to a better control on the SMBH centering in the host galaxy (see end of Section \ref{sec:thesimulatedclusters}). This numerical issue becomes particularly severe for cluster simulations, due to the frequent dynamical disturbances and mergers. For a discussion of different proposed {\it BH advection algorithms} see \cite{wurster2013}.}
It is also important to point out that we find even more stable results for the \rev{growth factors} of the BCGs at $z \lesssim 1.5$,  \rev{that is for their mass as a function of z normalized to the final value},
which is the main topic of the present paper.


\subsection{BCG mass evolution}
\label{sec:bcgevo}
The process of galaxy formation is affected by merger events, albeit their relative importance and nature have still to be fully clarified.
Therefore, in the study of the formation history of a galaxy, it is important to distinguish between the epoch at which its stars are born somewhere and the epoch at which they are assembled into a single galaxy. There are good theoretical reasons to believe that this distinction is particularly relevant for BCGs, as we briefly anticipated in Section \ref{sec:intro}. In order to keep the distinction as clear as possible, in the following we will refer to the former history as {\it creation} history and to the latter one as {\it assembly} history.

\subsubsection{Creation and Assembly Times}
\begin{figure}
	\includegraphics[width=\columnwidth]{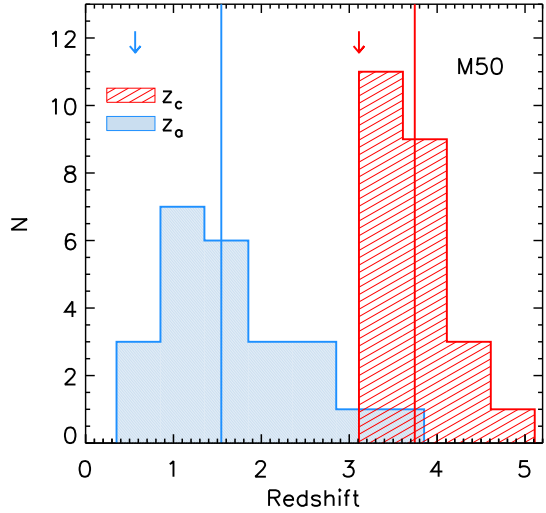}
    \caption{$z_a$ and $z_c$ distributions computed for M50 considering the Main Progenitor of the $z=0$ BCG.  Left and right vertical lines mark the medians of the $z_a$ and $z_c$ distribution respectively, while  arrows correspond to the medians for M10\%. For M30 the medians are 2.1 and 4 for assembly and creation respectively (not shown).}
    \label{fig:hist_zcre_zass}
\end{figure}

We define assembly redshift $z_{a}$ the redshift at which half of the final BCG stellar mass was assembled, and creation redshift $z_{c}$ the redshift at which half of its stellar mass was created.
The median values of the distributions of $z_{a}$ and $z_{c}$ for M30, M50 and M10\% are reported in Table \ref{tab:ztable}, both considering the BGC that resides in the main progenitor of the $z=0$ cluster as well as the main progenitor of the $z=0$ BCG. Note that the values of $z_c$ are by definition identical in both cases. For the BCG progenitor case, the distributions of $z_{a}$ and $z_{c}$ are shown in Fig. \ref{fig:hist_zcre_zass} for M50.

We find that both $z_a$ and $z_c$ depend on the aperture that was used to measure the mass. From Table \ref{tab:ztable} we see that the larger the aperture, the lower are both $z_a$ and $z_c$.
The former dependence suggests that the star particles in the outskirts are assembled later, as stated by the inside-out growth scenario \citep[e.g.][]{vandokkum2010,bai2014}, while the latter indicates that they also tend to be younger.

As for M50 Fig. \ref{fig:hist_zcre_zass} shows that half of the star particles that end up in the BCG at $z=0$ formed rapidly in a time interval $\sim 1.5 \mbox{Gyr}$. Indeed, these star particles were typically already formed by $z \sim 3.7$. Instead the assembly times $z_a$ have a much broader distribution spanning a range of $\sim 8 \mbox{Gyr}$ with a median assembly time at $z_a\sim1.5$.

\begin{figure}
	\includegraphics[width=\columnwidth]{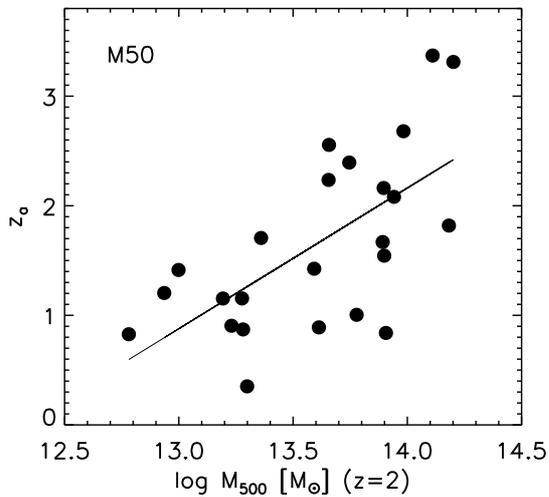}
    \caption{Assembly time $z_a$ as a function of cluster mass at z=2  considering the Main Progenitor of the $z=0$ BCG. The solid line is the best fit $y=1.28x  -15.82$. The Pearson correlation coefficient is 0.65.}
    \label{fig:za_vs_m500z2}
\end{figure}

In Fig. \ref{fig:za_vs_m500z2} we see a clear positive correlation with a Pearson coefficient of 0.65 between the assembly time and the cluster mass at  high z. Here in particular the cluster mass is calculated at $z=2$, but the correlation is still present down to $z\sim 1.3$, albeit weaker.
This can be understood taking into account that our complete sample of massive clusters features a relatively narrow range of masses of about 0.5 dex at $z=0$, while the corresponding range at $z=2$ covers about 1.5 dex. Therefore the clusters that were less massive at high redshift have grown much more up to $z=0$ than those that were more massive. The same has to be true for their BCG, provided that there is a link between the evolution of the BCG and its host cluster. This translates into a later (lower) BCG assembly time (redshift), defined as the time (redshift) at which half of the final mass was assembled.



\begin{table}
	\centering
	\caption{Medians of the assembly $z_a$ and creation $z_c$  distributions. Upper and lower bounds correspond to the 16\% and 84\% quantiles respectively.}
	\label{tab:ztable}
	
    \begin{tabular}{|l|l|r|l|r|l|r|}
		\hline
		\multirow{2}{*}{Dataset} &
          \multicolumn{2}{c}{M30} &
          \multicolumn{2}{c}{M50} &
          \multicolumn{2}{c|}{M10\%} \\
        & $z_a$ & $z_c$ & $z_a$ & $z_c$ & $z_a$ & $z_c$ \\
		\hline
        Prog. Clus. & $1.8\substack{0.3\\2.7}$ & $4.0\substack{3.5\\4.7}$ & $1.4\substack{0.4\\2.2}$ & $3.7\substack{3.3\\4.5}$ & $0.5\substack{0.1\\0.7}$ & $3.1\substack{2.9\\3.5}$ \\
        \\
        Prog. BCG   & $2.1\substack{1.4\\3.0}$ & $4.0\substack{3.5\\4.7}$ & $1.5\substack{0.9\\2.6}$ & $3.7\substack{3.3\\4.5}$ & $0.6\substack{0.3\\0.7}$ & $3.1\substack{2.9\\3.5}$ \\
	\end{tabular}
\end{table}

\subsubsection{Creation and Assembly Mass Histories}
\label{sec:evomass}
\begin{figure}
	\includegraphics[width=\columnwidth]{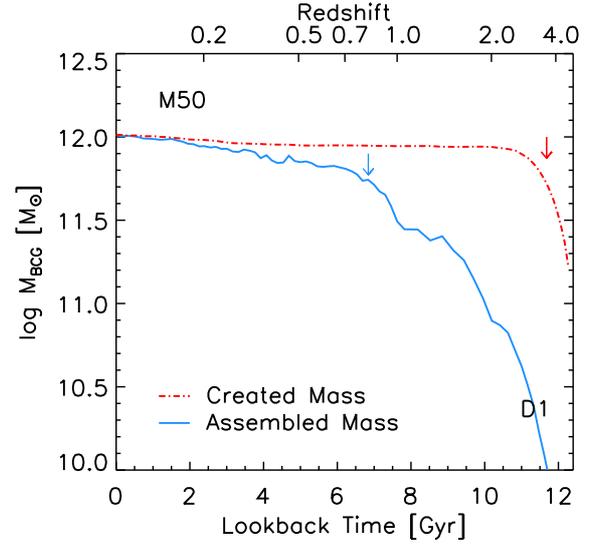}
    \caption{An example of the M50 main progenitor BCG evolution for one of the simulated galaxies. The assembled and created mass are represented by the solid and dashed line respectively. Downward arrows mark the assembly $z_{a}$ and creation $z_{c}$ redshift for each curve.
    }
    \label{fig:evoD1}
\end{figure}

\begin{figure*}
\includegraphics[width=7cm, height=7cm]{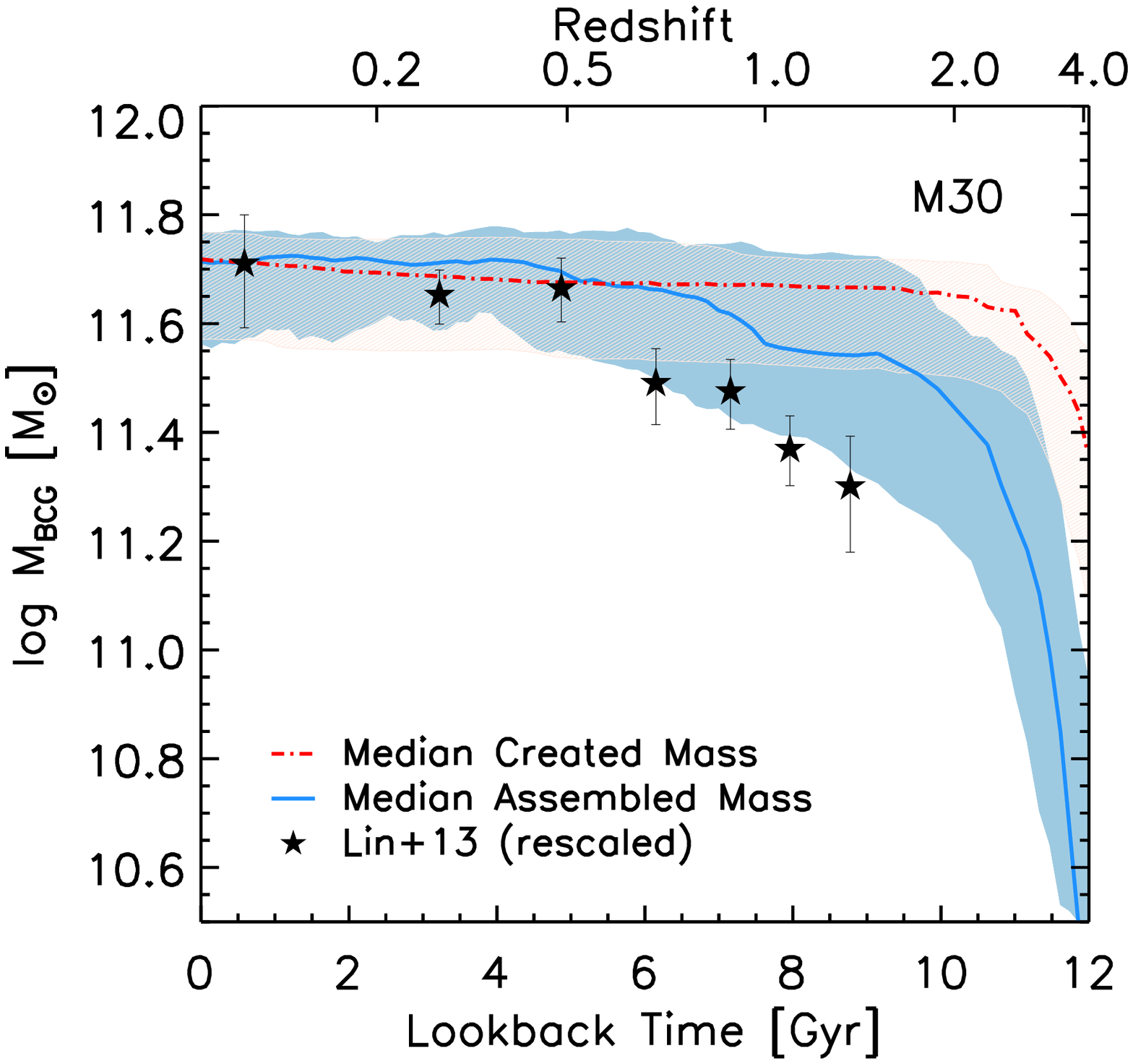}
\includegraphics[width=7cm, height=7cm]{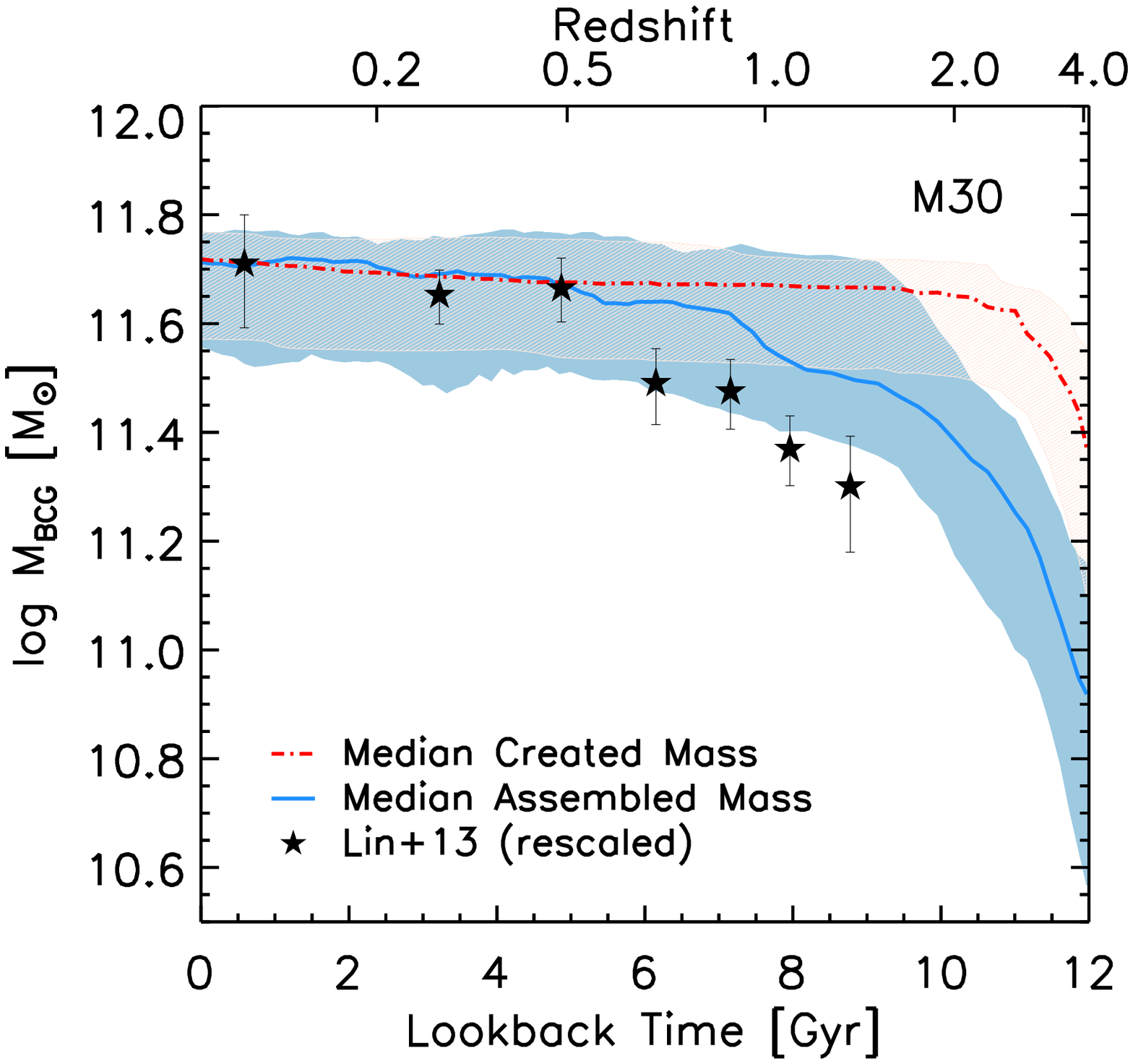}
\includegraphics[width=7cm, height=7cm]{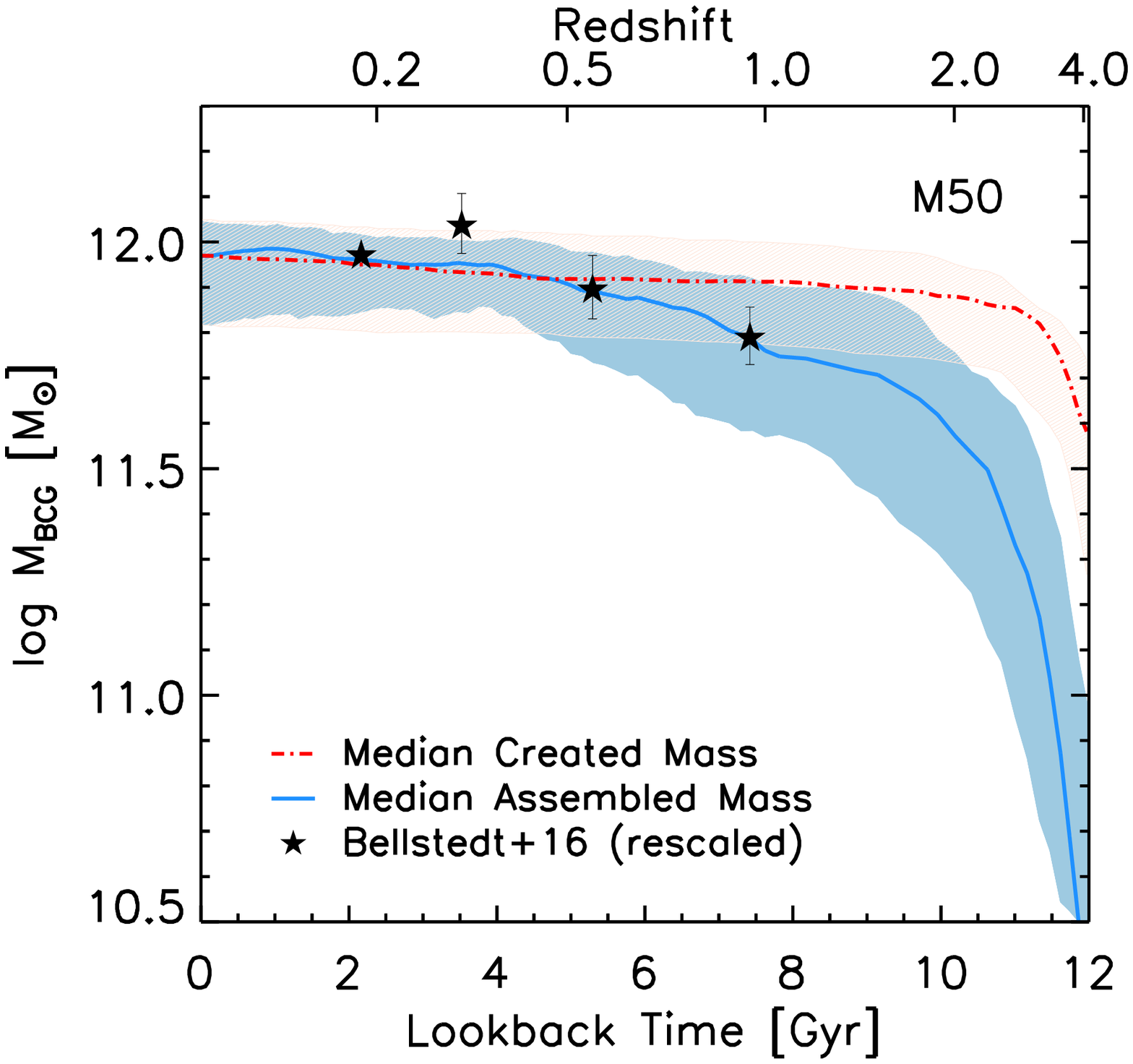}
\includegraphics[width=7cm, height=7cm]{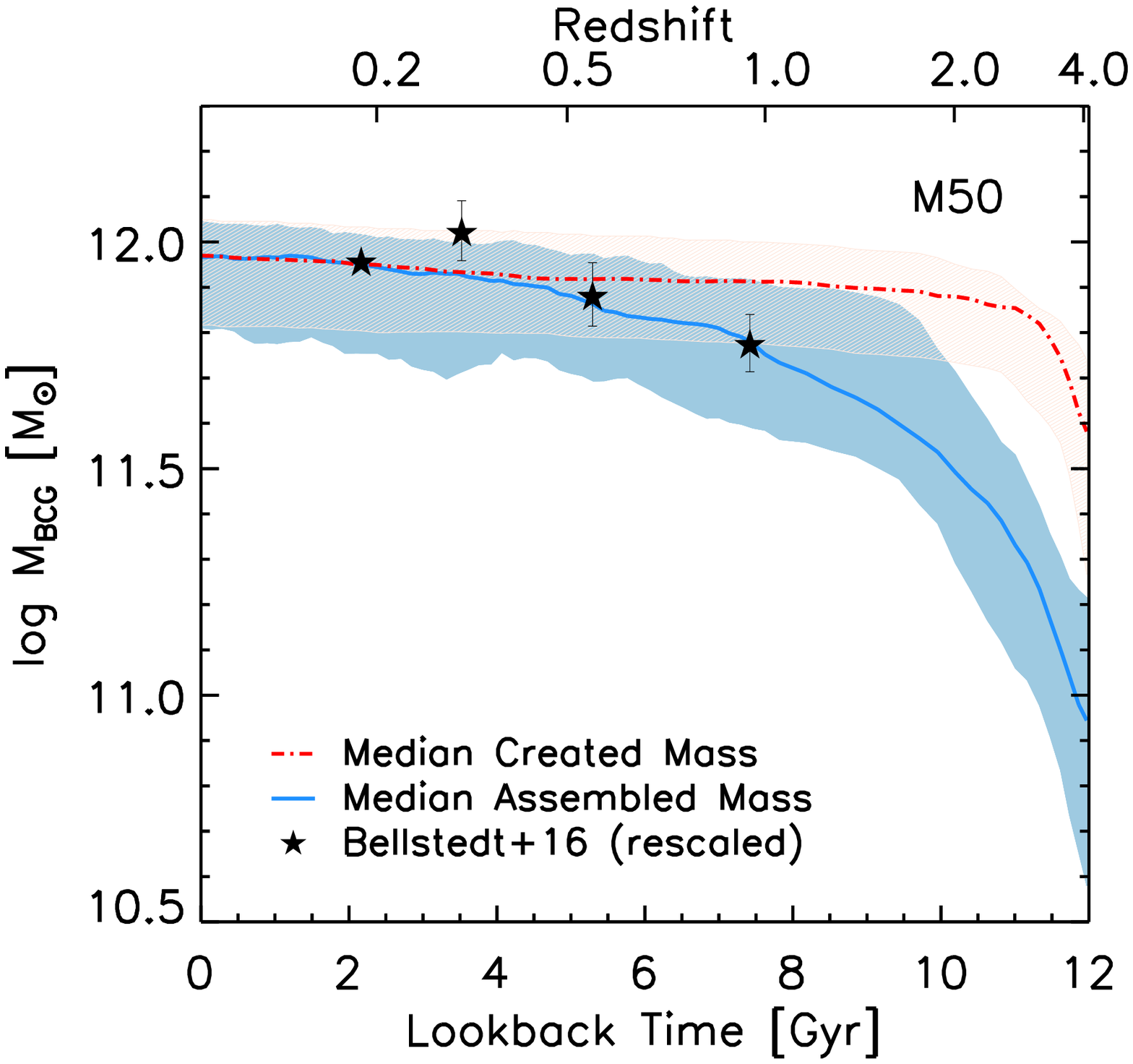}
\includegraphics[width=7cm, height=7cm]{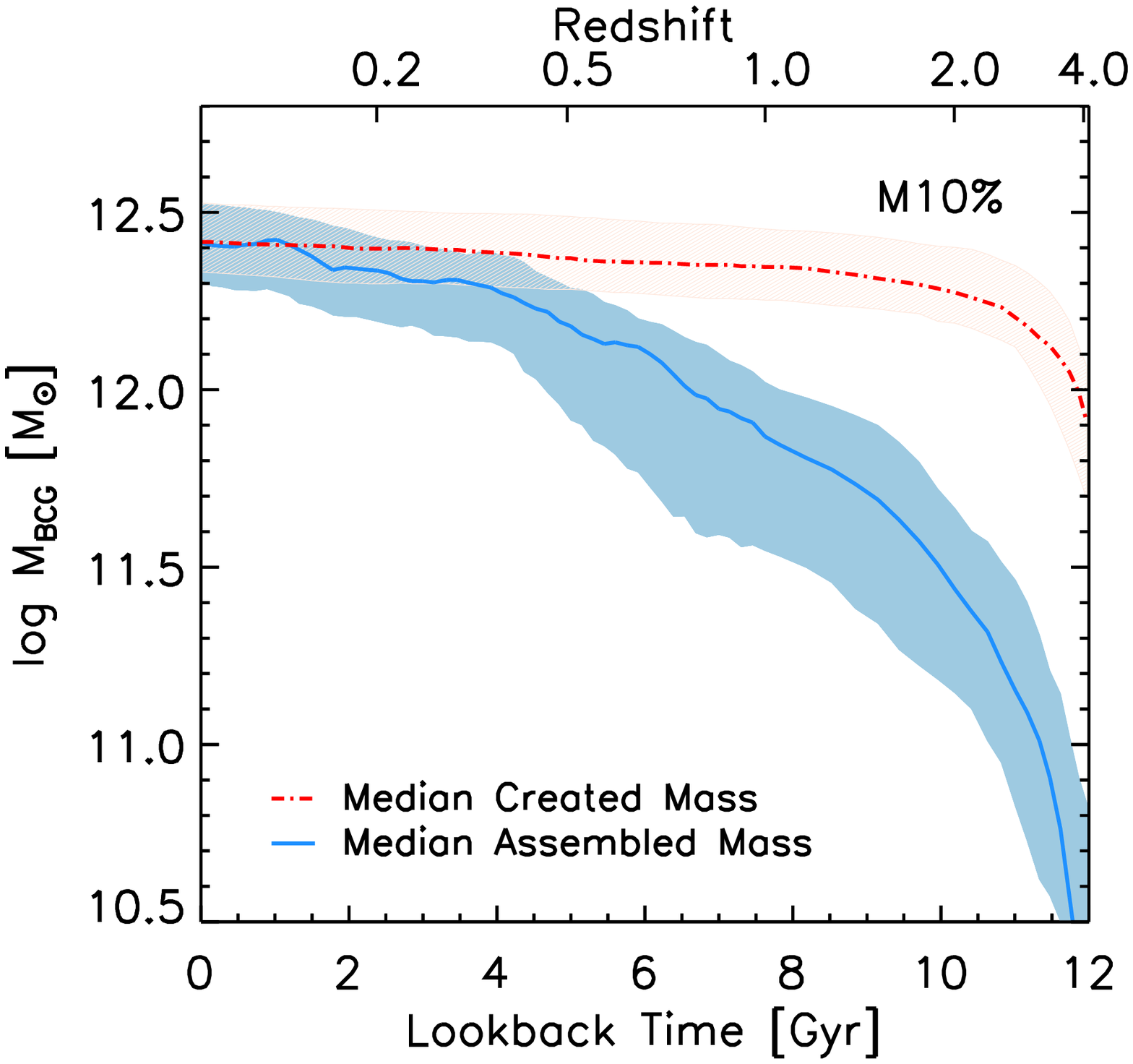}\includegraphics[width=7cm, height=7cm]{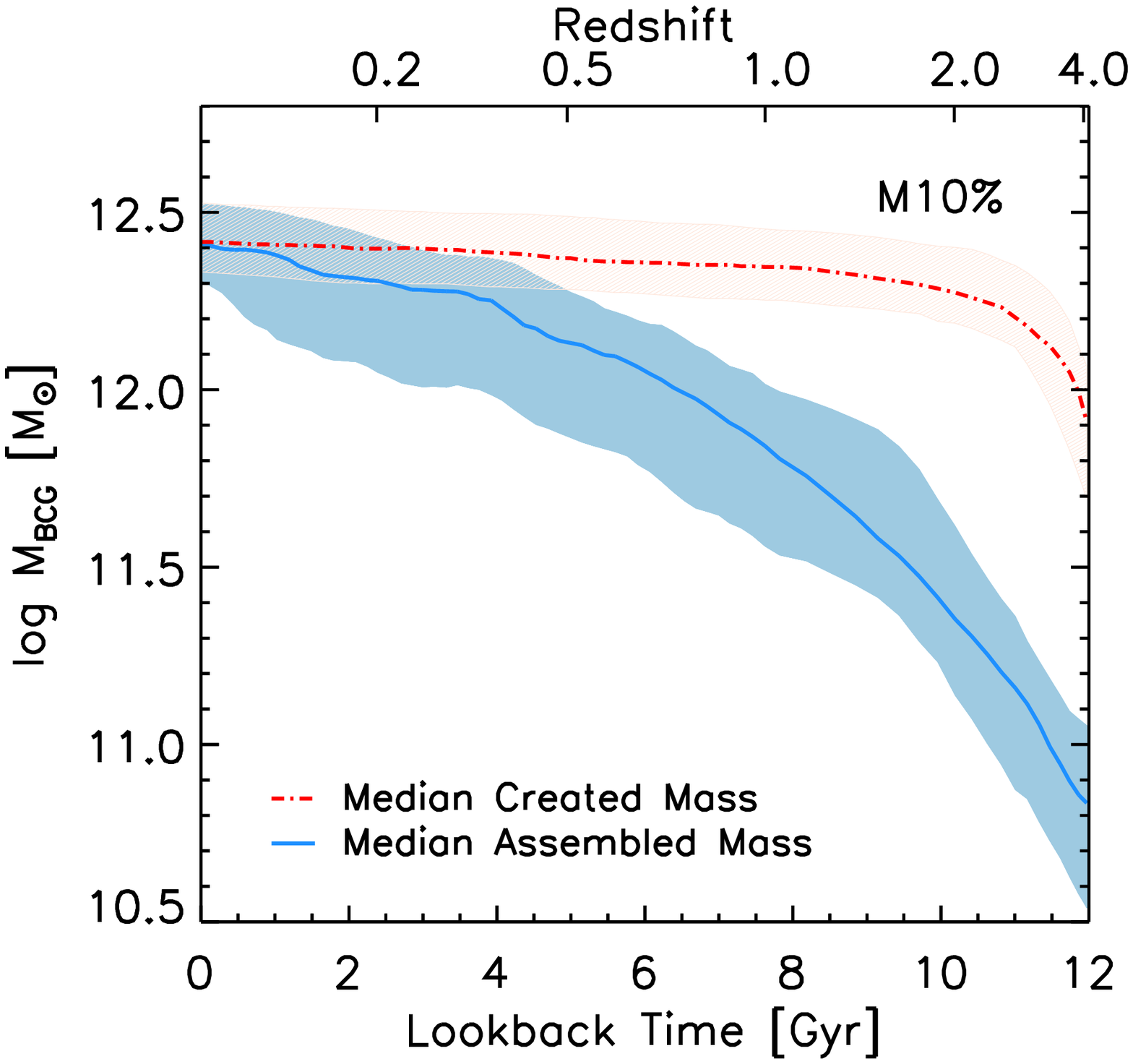}
\caption{Median mass evolution of the main progenitor galaxy of the BCGs (left column) and the BCGs in the main progenitor clusters (right column) for M30 (top), M50 (middle) and M10\% (bottom). The shaded regions enclose the 16\% and 84\% percentiles. Observational data is represented by stars. \citet{lin2013} data has been rescaled upward by a factor 1.23. See text for details.
 } \label{fig:masas_evo_fig}
\end{figure*}

\begin{figure}
\includegraphics[width=\columnwidth]{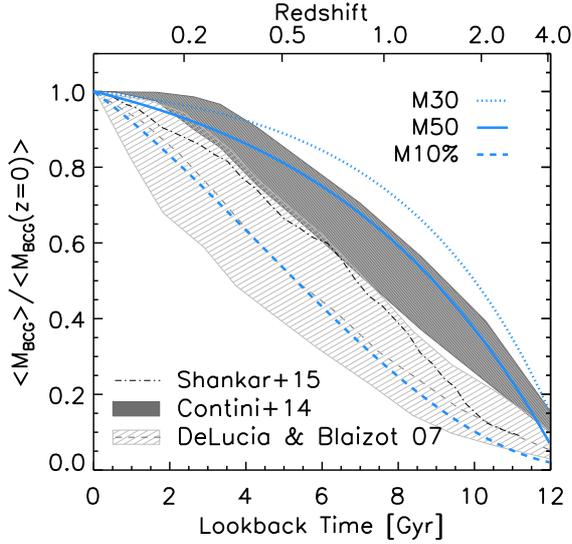}
    \caption{1/GMM($T$), which is  the evolution as a function of look-back time of the median assembled BCG mass in the main progenitors of the $z=0$ clusters normalized to the BCG mass at z=0. Dotted, solid and dashed thick lines correspond to the growth of M30, M50 and M10\% respectively in our simulations, as given by Equation \ref{eq:growfit}. The dark gray area \rev{covers the different evolutionary paths of the BCG mass (without ICL)} for the different ICL formation models presented in \citet{contini2014} SAM. Note however that they refer to a wider range of halo masses. Results from the \citet{shankar2015} semi-empirical model are plotted as a dash-dotted line, while those from \citet{delucia2007} SAM are in thin dashed line along with the 15\% to 85\% percentile range in light gray dashed area. In the latter  two models the considered halo mass range is similar to the one used in this work.
    }
    \label{fig:growz_fig}
\end{figure}

\begin{figure}
 	\includegraphics[width=\columnwidth]{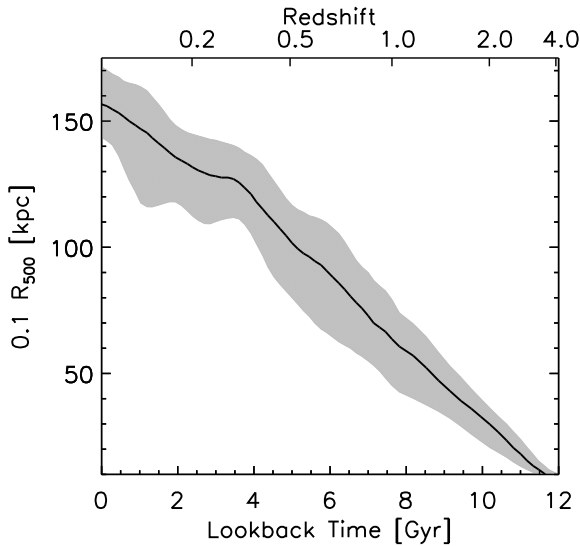}
    \caption{\rev{Median of the 0.1 $R_{500}$ radius of the main progenitor clusters as a function of time. M10\% is computed inside this radius. The 15\% to 85\% percentile range is comprised within the gray region.}}
    \label{fig:grow01r500_fig}
\end{figure}

\begin{figure}
 	\includegraphics[width=\columnwidth]{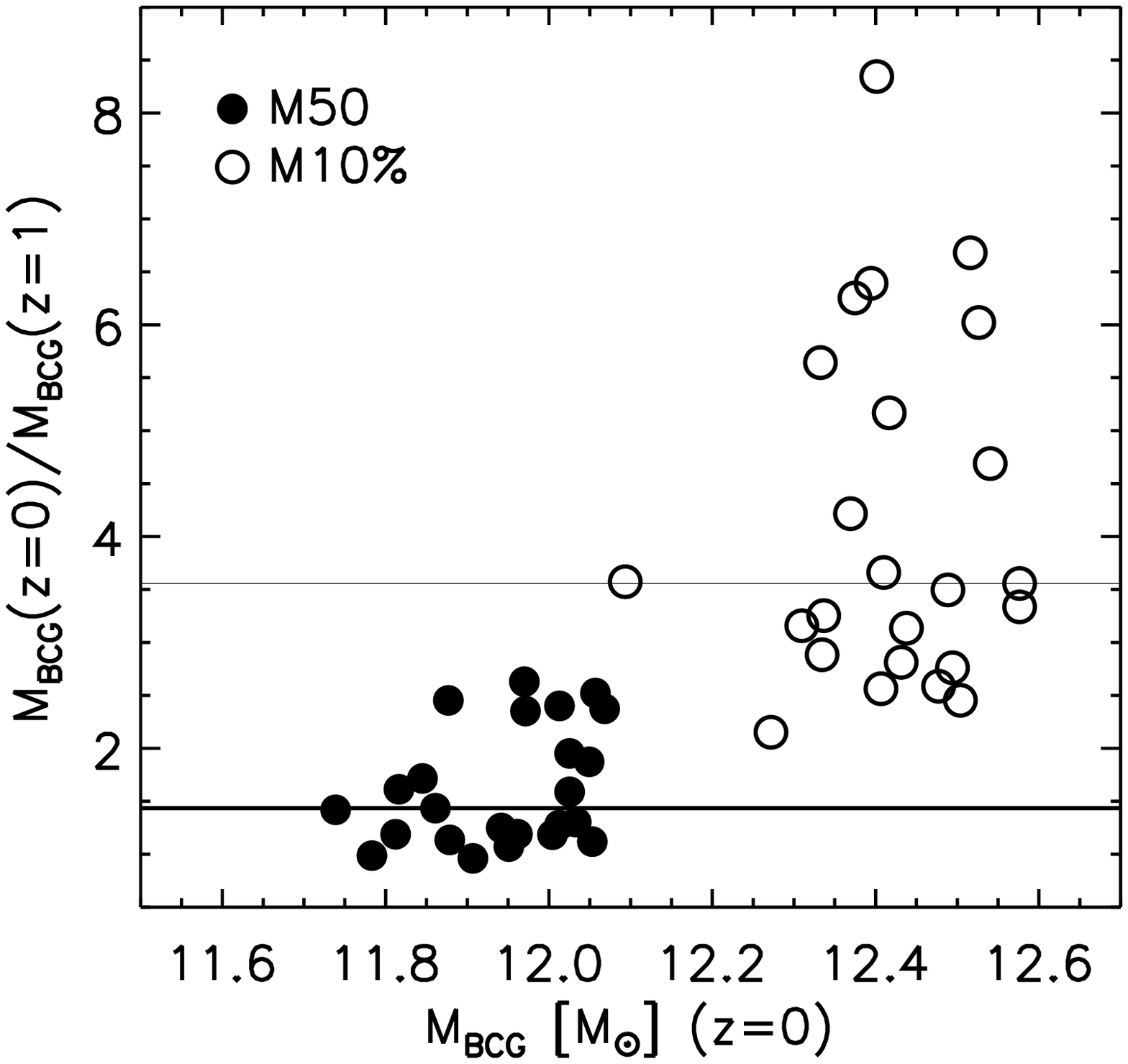}
    \caption{The individual BCG mass growth factor between z=1 and z=0 as a function of BCG mass. Only the sample of BCGs extracted from main progenitor clusters is considered. Horizontal thick and thin lines stand for the median growth factor MGF, 1.4 and 3.6, for M50 an M10\% respectively. The median growth factor for M30 is $\sim$ 1.2 (not shown)}
    \label{fig:grow1_fig}
\end{figure}

Fig. \ref{fig:evoD1} shows the evolution of both the assembled and created M50 mass for one of our simulated BCGs (solid and dashed line respectively). In this particular case half of the stellar mass that ends up at redshift zero within the central 50 kpc, has already been created at $z_{c} \sim 3$  (right arrow), while the assembly of half of the mass occurs at $z_{a}\sim$ 0.8 (left arrow).

Information  similar to that shown in Fig. \ref{fig:evoD1}, but in a statistical sense for the whole cluster sample, is shown in Fig. \ref{fig:masas_evo_fig}. In this case we show the median assembled and created mass history (blue solid and red dashed lines respectively) of both the main progenitor of the $z=0$ BCGs (left panels), and of the BGC residing in the main progenitor of the $z=0$ cluster (right panels), as discussed in Section \ref{subsec:mbcgmclus}. The shaded regions enclose the 16\% and 84\% percentiles of the mass distribution at each redshift.

Complementary information is given in Table \ref{tab:growtable} where we report growth factors of M30, M50 and M10\% for our sample of BGCs from $z=2$ and $z=1$ to $z=0$. They are computed in different ways. As already mentioned, we can follow either the main progenitor of the final cluster (first and third rows), or the BCG in the main progenitor of the final BCG (second and fourth rows). In addition, we can compute either the median of the growth factors for individual BCG (columns with header MGF; see also Fig. \ref{fig:grow1_fig}) or the growth factor of the median mass at each redshift (columns GMM; obtained from Fig. \ref{fig:masas_evo_fig}). Note that the quantities that should be compared to estimates from observational data are the GMM of the main progenitor cluster, reported in boldface in the table. This is because observational analyses usually compute the ratio between the typical masses of BCGs residing in cluster samples selected around two different redshifts. Moreover, at least in the optimum cases, the cluster masses are chosen at different redshifts in order to approximate as much as possible an evolutionary sequence \citep[e.g.][]{lin2013,bellstedt2016}.

The first point to notice in Fig. \ref{fig:masas_evo_fig} is that the  mass evolution histories represented in the left and right panels are almost identical within a few percent at least up to $z \simeq 1$ (that is the maximum redshift at which most observed BCG mass growth factors are reported), and very similar even at higher redshift.
From top to bottom we show the evolution of M30, M50 and M10\%. It is apparent that along this sequence the assembled mean mass growth (blue solid curve) becomes more and more pronounced. This is partly due to the fact that while the galaxies become more massive and consequently larger, an increasing fraction of their mass is excluded by smaller apertures.
Conversely, the created mass growth is very similar in the three cases, indicating that only a mild stellar age gradient is present in the central cluster regions in our simulations.

In the top row M30 evolution is compared to the observational results by \cite{lin2013}, who studied the stellar mass growth of BCGs in the IRAC shallow cluster. As mentioned in Section \ref{sec:clussel} observational data should be confronted with curves in the right panels, nevertheless we plot them in both panels for sake of comparison. \cite{lin2013} estimated the BCG mass within a redshift-independent aperture of 32 kpc (physical). As such, their masses can be in principle directly compared to our evolution of M30.
At different redshifts they used samples of clusters that have been selected in such a way to mimic an evolutionary sequence. In order to facilitate the comparison between data and simulation mass growth, we rescaled \cite{lin2013} BCG masses upward by a factor 1.23 to match their lowest redshift bin with our assembled mass curve.
Note that their typical cluster mass is about a factor four smaller than ours, which justifies their somewhat smaller median BCG mass at low redshift.
For the same reason it is conceivable that going down in redshift, the 30 kpc apertures exclude a mass fraction that increases faster in our sample than in the observed one. In this sense, it could be more appropriate to compare the relative growth measured by \cite{lin2013} with that predicted by our simulations within a somewhat larger aperture, such as M50, shown in the middle row of Fig. \ref{fig:masas_evo_fig}. Independently of these details, a reasonable agreement between simulated and observed mass assembly is apparent from the figure, in particular concerning  the very limited growth at $z \lesssim 0.5$.
Also \cite{zhang2016} estimated a growth factor for the BCG masses within 32 kpc; they found this factor to be 1.35, very similar to our median value of 1.3 within 30 kpc (see Table \ref{tab:growtable} under the GMM label). However, the value reported by \cite{zhang2016} refers to significantly less massive clusters.

More recently \cite{bellstedt2016} compared masses of BCGs in four redshift bins from 0.18 to 0.95. The subsamples of clusters in each bin were selected to have matching mass distribution, extrapolated to $z=0$ by means of a growth history derived from simulations, with respect to one {\it reference} bin. They used as reference either the bin centered around $z=0.18$ or $z=0.35$. This choice was done again in the effort of comparing BCGs in clusters that can be regarded as belonging to an evolutionary sequence. Their determinations of the mass ratio between three higher redshift bins and that centered at 0.18 are compared to the growth of M50 in the middle row of Fig. \ref{fig:masas_evo_fig}. We use the 50 kpc aperture in this case because \cite{bellstedt2016} estimated masses from luminosities computed with {\tt MAG\_AUTO}, which according to \cite{stott2010} are well approximated by 50 kpc fluxes for BCGs. The observational estimates have been rescaled to the median mass of our simulated sample at $z=0.18$.   \cite{bellstedt2016} did the same when comparing to other models and observations. Our simulations reproduce very well the mild mass increase of BCGs below $z\simeq 1$ suggested by these observational results.

\begin{table*}
	\centering
	\caption{BCG mass growth factors. MGF: Median of the Growth Factors distributions \rev{with its 16\% and 84\% percentiles}. GMM: Growth of the Median Mass, \rev{errors are determined by bootstrap re-sampling}. The numbers which should be compared to estimates coming from observational samples at different redshifts, properly selected, are the GMM for the main progenitor clusters. They are printed in boldface.}
	\begin{tabular}{clcccccc} 
		\hline
        \multirow{2}{*}{$\Delta z$} & \multirow{2}{*}{Dataset} &
          \multicolumn{2}{c}{M30} &
          \multicolumn{2}{c}{M50} &
          \multicolumn{2}{c|}{M10\%} \\
        & & MGF & GMM & MGF & GMM & MGF & GMM \\
        \hline
        \multirow{2}{*}{0 - 1} &
          Progenitor Cluster & $1.2\substack{0.97\\1.9}$ & ${\bf 1.3\pm0.3}$ & $1.4\substack{1.1\\2.4}$ & ${\bf 1.6\pm0.2}$ & $3.6\substack{2.6\\6.3}$ & ${\bf 3.6\pm0.6 }$\\ \\
          &Progenitor BCG     & $1.3\substack{1.0\\1.9}$ &   $1.4\pm0.2$    & $1.6\substack{1.1\\2.5}$ &  $1.6\pm0.3$     & $4.9\substack{2.6\\7.4}$& $3.5\pm1.0$\\
        \hline
		\multirow{2}{*}{0 - 2} &
          Progenitor Cluster & $2.2\substack{1.5\\3.2}$ & ${\bf 2.1\pm0.3}$ & $2.7\substack{1.8\\4.5}$ & ${\bf 2.9\pm0.5}$ & $11.7\substack{6.5\\21.4}$ & {$\bf 11.0\pm2.0$}\\
          \\
          &Progenitor BCG   & $2.3\substack{1.3\\7.0}$ &    $1.9\pm0.2$       & $3.5\substack{1.7\\12.8}$ &   $2.5\pm0.4$        & $12.9\substack{7.4\\30.9}$ &$9.5\pm2.0$\\
        \hline
	\end{tabular}
	\label{tab:growtable}
\end{table*}

At late times $z \lesssim 0.5$ our simulated BCGs are characterized by very little mass growth, in agreement with some observational works \citep[e.g.][]{lin2013,oliva2014,inagaki2015}.

\rev{We remark also that some estimates of the BCG mass growth via mergers, based on counting companion galaxies that are likely to merge with the BCG by the present time, point to a moderate growth by means of this growth channel for $z \lesssim 0.8$ \citep[e.g.][]{liu2009,burke2015}.}

The GMM factors in the redshift range $z=0$ to 4 obtained for BCGs residing in the progenitor clusters can be obtained from the analytic representations given by
\begin{eqnarray}
\label{eq:growfit}
& & 1/\mbox{GMM}_{30}(T) =  1.045 -0.045 \times 1.283^T \nonumber \\
& & 1/\mbox{GMM}_{50}(T) =  1.145 -0.145 \times 1.182^T\\
& &1/\mbox{GMM}_{10\%}(T) = 1-0.0740 \times  T  - 0.00628 \times T^2 +\nonumber \\ 	
& & \qquad \qquad \qquad   + 0.00047 \times T^3\nonumber
\end{eqnarray}
where $T$ is the lookback time in Gyr. The corresponding mass evolutions normalized to the BCG mass at z=0 are shown in Figure \ref{fig:growz_fig}.

In this figure, the mass growth found in our simulated BCGs is also compared with those published for a few semi-analytical and semi-empirical models. While the growth of ``total" BCG mass predicted by \citet{delucia2007} SAM is much stronger than that of M50 or M30, it closely resembles that of M10\%\footnote{strictly speaking the result by \cite{delucia2007} refers to the assembly the main progenitor of the BCG, while our curve refers to the BCGs in the main progenitor of the cluster, for a more proper comparison with observed samples. However at least in our case the two choices lead to very similar results.}. In particular, the median M10\% growth from $z=1$ to $z=0$ turns out to be 3.5 or 3.6 (Table \ref{tab:growtable} under the GMM label), almost identical to that reported in the mentioned work.
Note that the results of this SAM refers to the high mass end of the cluster mass function, like our simulations. \rev{However, the similarity of the relatively fast growth should not be over-interpreted, in that it is due to entirely different reasons. In the former case it is largely driven by the time increase of the radius within which we are computing masses.   This radius, shown in Fig.\ \ref{fig:grow01r500_fig}, is large enough to comprise all the galaxy light at low redshift, still without substantial contribution from the ICL, while excluding an increasing fraction of the galaxy at higher and higher redshift.
To assess this, we have computed surface brightness maps of the BCG region of our simulated clusters at redshift 0, 1 and 2. We used the same procedure described  by \cite{cui2014} when studying the ICL component in a previous version of our simulations.
The radius at which our simulated BCGs drop to a rest frame surface brightness of $\mu_B \sim 25 {\rm~mag~arcsec^{-2}}$ (or $\mu_V \sim 24{\rm~mag~arcsec^{-2}}$) is on average 130, 110, and 100 kpc at redshift 0, 1 and 2 respectively. This brightness threshold is a classical operational value to define the galaxy limit \citep{devaucou1991}, which tends to underestimate the real extension \citep[][and references therein]{feldmeier2004}. An inspection to Fig.\ \ref{fig:grow01r500_fig} thus confirms that  M10\% can be regarded as a reasonable proxy of the ``total'' simulated BCG mass in the local universe, but
progressively underestimates it at higher z.
Moreover, we verified that the fraction of M10\% contributed by regions fainter than $\mu_V \sim 25{\rm~mag~arcsec^{-2}}$ never exceeds a few percent at $z=0$, and it is totally negligible at $z=1$ and 2. The latter surface brightness limit is close or more often somewhat brighter than the typical thresholds used to identify the ICL both in observations \citep[e.g.][]{feldmeier2004,zibetti2005,krick2006,krick2007,montes2018} as well as in simulations \citep[e.g.][]{rudick2011,cui2014}.
Thus in our simulations $0.1 R_{500}$ never includes any important ICL contribution.
On the other hand, the fast growth of the total BCG mass predicted by the \cite{delucia2007} SAM is likely related to the absence of any  treatment for the late development of the ICL, which once included slows down the late growth of the BCG component.
Model computations agree that the ICL component begins to be important at $z \lesssim 1$ \citep[see also][]{monaco2006,murante2007}. In particular}
\cite{contini2014} introduced in the same SAM used by \cite{delucia2007} several possible prescriptions for the formation of the ICL component.
The dark gray region in Fig. \ref{fig:growz_fig} \rev{covers the various evolutionary paths of the BCG mass (without ICL) for all the different models for the ICL formation} considered by \cite{contini2014}. We remark that their results are averaged over a range of halo masses extending to significantly lower values than ours. In any case, it can be seen that the inclusion of the ICL makes the mass evolution of BCGs much weaker.  As such, the BCG mass growth turns out to be much closer to both the growth implied by the data reported in Fig. \ref{fig:masas_evo_fig}, and the growth predicted by our simulations within fixed physical apertures.



We also show in the figure the BCG mass evolution predicted by the semi-empirical sub-halo abundance model put forward by \cite{shankar2014,shankar2015}. In this class of computations most galaxy properties along the merger tree of the central galaxy are constrained by means of empirical relationships. As for the central galaxy, this is done until a major merger occurs. The purpose is to elucidate the role of mergers in determining the size and mass evolution of central galaxies, under different assumptions on their modality. The reference model by \cite{shankar2015}, which we plot for the same halo mass range of our simulated BCGs (Shankar 2018, private communication), features an evolution in between that of our M50 and M10\%. \cite{shankar2015} favored a model without stripping and with parabolic orbits for satellite mergers, in order to maximize the predicted size evolution. With these choices, the mass evolution is maximized as well.

In conclusion the strong dependence of the assembly history on the adopted mass definition highlights that comparisons between SAM and observations are not straightforward and should be done with proper care.

The individual M50 and M10\% mass gains between $z=0$ and $z=1$ for all the simulated BCGs in the main progenitor clusters are plotted in Fig. \ref{fig:grow1_fig} as a function of BCG mass. The medians of these growth factors along with those corresponding to the redshift range 0-2 are reported in Table \ref{tab:growtable} under the MGF labels. No significant differences with respect to the GMM factors mentioned above is evident. We note a mild positive correlation of the mass growth with the mass of the BCG when considering M30 and M50 with Pearson correlation coefficients 0.35 and 0.33 respectively.


To the best of our knowledge, there are no specific studies based on hydrodynamical simulations devoted to investigate the mass growth of BCGs. However also \cite{martizzi2016} reported a moderate mass growth since $z=1$ in their AMR simulations of 10 galaxy clusters with $M_{vir} \simeq 10^{14} \mbox{M}_\odot$, with a growth factor $\lesssim 2$ within 50 kpc.

\begin{figure}
\includegraphics[width=\columnwidth]{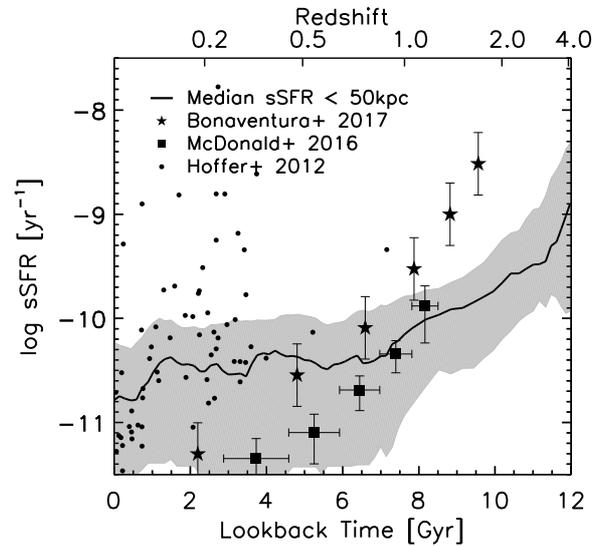}
\caption{Evolution in time of the median specific star formation rate inside 50kpc for BCGs in the main progenitor clusters, compared to observational results. The shaded area enclose the 16\% and 84\% percentiles.
 } \label{fig:sfr_evo_fig}
\end{figure}

\begin{figure}
    \includegraphics[width=\columnwidth,trim=0cm 9cm 9cm 0cm,clip=true]{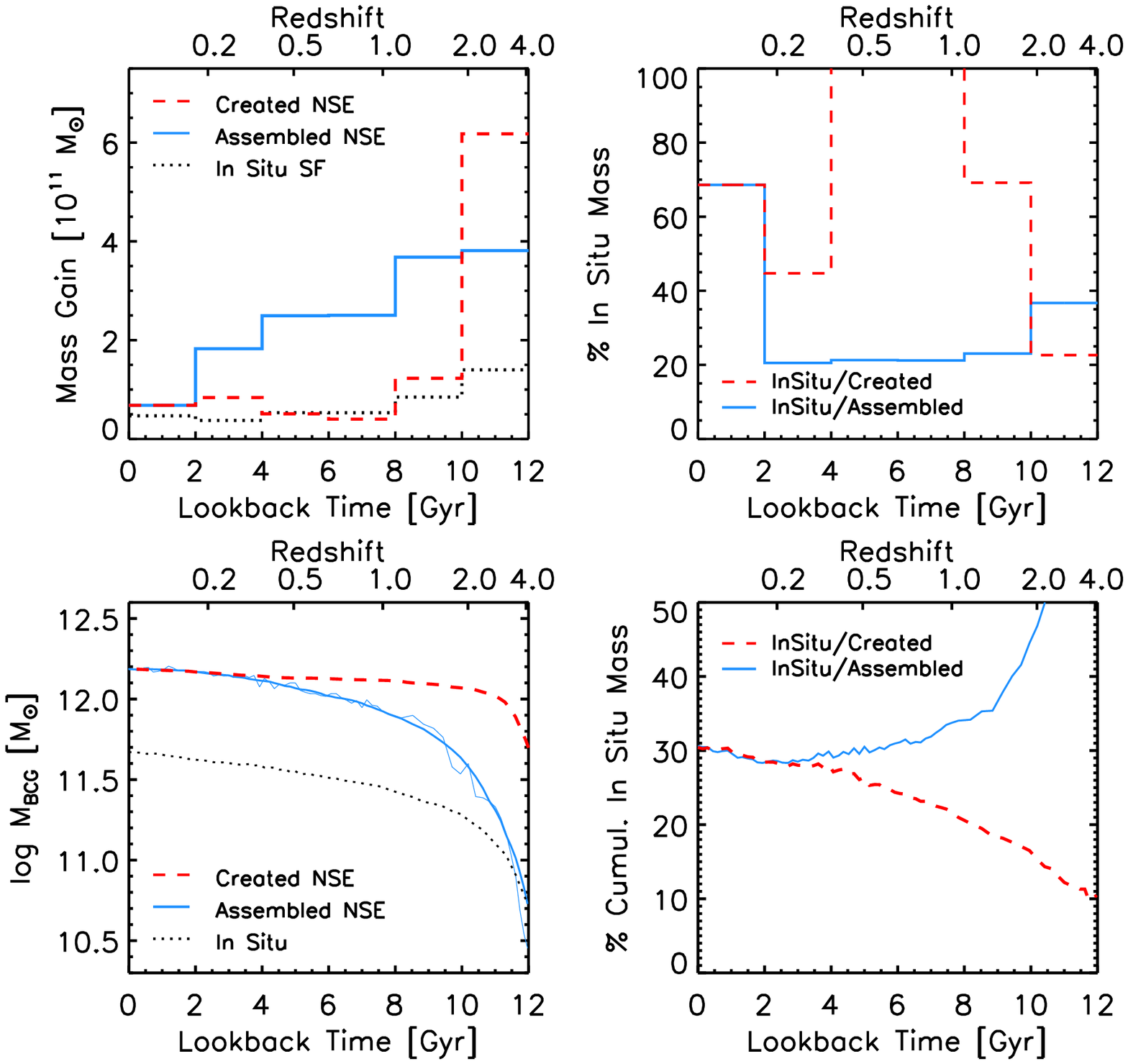}
    \caption{Gained masses per 3 Gyr bin. The dashed and solid lines represent the median created and assembled M50 stellar mass respectively, without considering stellar evolution. Dotted line stands for the stellar mass formed by in situ star formation.}
    \label{fig:insitu}
\end{figure}

\subsection{{\it In Situ} and {\it Ex Situ} Star Formation}
The star formation histories of the simulated BCGs feature typically a fluctuating behavior around $\sim 30 \,\mbox{M}_\odot \mbox{yr}^{-1}$ from $z\simeq 0.1$ to $z \simeq 1$, with hints of a decrease over the last Gyr and a clear increase above $z \sim 1$. Fig. \ref{fig:sfr_evo_fig} shows the evolution of the median specific star formation of the BCGs in the main progenitor of the $z=0$ cluster. This is compared to the recent observational estimates of \cite{bonaventura2017}, \cite{mcdonald2016} and \cite{hoffer2012} which include IR data. These observational estimates show some level of disagreement among them, well above the quoted errors. Nevertheless, they seem to indicate that simulated BCGs lack intense enough SF at high redshift, while possibly exhibiting an excess of residual SF at low redshift. The first point is reminiscent of what we already noticed in \cite{granato2015}, when comparing with the observed Herschel fluxes of candidate high-redshift protoclusters. However in that case the focus was on a much more extended region than just the BCG progenitor.

It is worth studying the contribution of this {\it in situ} star formation to the total assembled and created mass.
Fig. \ref{fig:insitu} compares the median M50 gain due to SF occurring in the main progenitor of the BCG ({\it in situ} SF, black dotted line) with the median global mass gain (``Assembled", blue solid line) and with the median mass increase of all the star particles that end up in the BCGs at $z=0$ (``Created", red dashed line). All these rates are computed in bins of 2 Gyr in look-back time. For the sake of a proper comparison with the {\it in situ} SF, the latter two are computed using the initial mass of the stellar particles, that is ignoring the mass loss due to stellar evolution (NSE meaning No Stellar Evolution).
At first sight it could seem surprising that sometimes the {\it in situ} star formation slightly exceeds the created mass gain. However it should be kept in mind that the latter refers to the star formation of the particles that end up in the inner 50kpc of the BCGs. Depending on the dynamical history, a certain fraction of the stellar particles formed {\it in situ} can leave later the very central region of the cluster.


By inspecting the solid blue line in Fig. \ref{fig:insitu}, it can be recognized that from $z=4$ to $z \gtrsim 0.2$ about 20\% to 30\% of the increase of the assembled mass is due to {\it in situ} SF. 
The comparison of the other two lines tells us  that the {\it in situ} SF can account for an important fraction of the stars that were born in the last 10 Gyr and ended up in the $z=0$ BCG. On the other hand, at earlier times, when all the rates were higher, the contribution of the {\it in situ} SF was much less important. Actually at high redshift there are often other star forming progenitors having similar or even higher masses than that in the main progenitor branch.

By the end of the evolution the {\it in situ} SF produced typically an amount of mass close to 1/3 of the M50 mass of BCGs.
Between $z=2$ and $z=1$ we find a median growth due to {\it in situ} star formation of $\sim 25\%$, in good agreement with the observational estimate of $\sim 20\%$ reported by \cite{zhao2017}.

\section{Summary}
\label{sec:summary}
This work was devoted to analyze the mass growth of BCGs that form in a set of 24 zoom-in simulations of massive galaxy clusters with $M_{500}> 7\times 10^{14} \mbox{M}_\odot$. These simulations include sub-resolution description of star formation, stellar feedback and AGN feedback. Our results turns out to be in broad agreement with most recent observational results and can be summarized as follows:
\begin{itemize}
\item
The final masses of our simulated BCGs as a function of cluster mass are in reasonable agreement with observations. They turn out to be smaller by a factor 2 to 3 than those predicted by recent state of the art galaxy cluster simulations \citep{bahe2017,pillepich2017}. A few tests we performed at 3 times higher mass resolution resulted into final BCG masses differing by less than \rev{10\%} from our standard resolution runs, \rev{without any systematic trend for a mass increase or decrease at higher resolution.}
\item
We have constructed mass growth histories up to $z=4$ for the stellar mass within 30 kpc and 50 kpc, and within 10\% of the $R_{500}$. For this purpose, we followed back in time both the main progenitor of the final BCGs, and the BCG residing in the main progenitor of the $z=0$ clusters. In principle the latter choice is more suitable for comparison with estimates of the growth factors coming from observations. However we found only minor differences at least up to $z\sim 1.5$, that is in the redshift regime to which most data refer. Analytic prescriptions for the growth histories are given in Equation \ref{eq:growfit}.
\item
In concordance with some observational works \citep[e.g.][]{lin2013,oliva2014,inagaki2015} we find very little growth for the stellar mass (within 30 and 50 kpc) up to $z \sim 0.5$.
Up to $z=1$ the growth factors we find increase with the aperture. Our median mass growths by a factor 1.3 and 1.6 for M30 and M50 respectively are in good agreement with the most recent observations adopting similar or equivalent apertures \citep[e.g.][]{lin2013, zhang2016, bellstedt2016}.
\item
These growth factors are significantly smaller than those found in the SAM by \cite{delucia2007}, which turns out to be similar to the one we find for the stellar mass within a fraction of 10\% of the $R_{500}$. That SAM  describes the evolution of the ``global'' BCG stellar mass and moreover does not include any treatment of the development of the ICL. \rev{The relatively fast mass growth within $0.1 R_{500}$ is instead mostly driven by time increase of this radius (Figure \ref{fig:grow01r500_fig})}.
The strong dependence of the assembly history on the adopted mass definition highlights that comparisons between SAM and observations are not straightforward and should be done with proper care.
\item
Half of the star particles that end up in the inner 50 kpc of the BCGs was typically already formed by redshift $\sim$ 3.7. On the contrary the assembly of half of the stellar mass of the BCG occurs on average at lower redshifts $\sim 1.5$. This assembly redshift shows a positive correlation with the mass that the cluster attained at high $z \gtrsim 1.3$, and an anti-correlation with aperture.
\item
The typical formation history of the BCG can be broadly divided into two phases. At $z \gtrsim 2$, the contribution of the {\it in-situ} (i.e. main progenitor) SF is minor compared to the global SF occurring in all the progenitors that end up into the local BCG. In this high redshift regime other progenitors compete and may surpass in mass the main one. On the contrary, going to lower redshift when the global SF rapidly decreases by about one order of magnitude the {\it in-situ} star formation accounts for more than half of the total. \rev{Nonetheless, in this phase, accretion of stars formed elsewhere is the main channel of the BCG mass gain (except in the last $\sim 2$ Gyr)}. By the end of the evolution the {\it in situ} SF produced typically an amount of mass close to 1/3 of the M50 mass.
Modulo the still important observational uncertainties, simulated BCGs could lack intense enough SF at high redshift, while possibly exhibit an excess of residual SF at low redshift.

\end{itemize}

It is interesting to note that both the decrease of the assembly redshifts and the increase of the mass growth factors with the aperture are compatible with the {\it inside-out} scenario. This suggests that most of the late mass growth of BCGs occurs typically by means of minor mergers or diffuse accretion rather than major mergers.
\rev{The contribution of minor/major mergers to the late mass growth of BCGs is still a debated issue in observational works using the pair fraction as a proxy for merger rates \cite[e.g.][]{mcintosh2008,liu2009,edwards2012,burke2013,burke2015,groenewald2017}}. We plan to analyze in detail the contribution of the different growth channels \rev{both to the BCG and ICL growth} with a future higher resolution set of simulations.

\section*{Acknowledgements}
We are indebted with the anonymous referee for several suggestion that improved the quality of our paper.
We thank Francesco Shankar for useful discussions and for promptly providing us  model results in electronic form.
GLG and GM thank IATE, Argentina for warm and extended hospitality during the development of the present work.
This project has received funding from the Agencia Nacional de Promoci\'on Cient\'ifica y Tecnol\'ogica de la Rep\'ublica Argentina under the PICT 2417-2013 grant, from the Consejo Nacional de
Investigaciones Cient\'ificas y T\'ecnicas de la Rep\'ublica Argentina
(CONICET), from the Secretar\'ia de Ciencia y T\'ecnica de la Universidad Nacional de C\'ordoba - Argentina (SeCyT) and from
the European Union's Horizon 2020 Research and Innovation Programme under the Marie Sklodowska-Curie grant agreement No 734374. We acknowledge also financial support from PRIN-MIUR 2015W7KAWC, the INFN INDARK grant, Consorzio per la Fisica of Trieste. SP is ``Juan de la Cierva'' fellow (ref. IJCI-2015-26656) of the {\it Spanish Ministerio de Econom{\'i}a y Competitividad} (MINECO) and acknowledges additional support from the MINECO through the grant AYA2016-77237-C3-3-P and the Generalitat Valenciana (grant GVACOMP2015-227).
Simulations have been carried out using MENDIETA Cluster from CCAD-UNC, which is part of SNCAD-MinCyT (Argentina) and MARCONI at CINECA (Italy), with CPU time assigned through grants ISCRA C, and through INAF-CINECA and University of Trieste - CINECA agreements. The post-processing has been performed using the PICO HPC cluster at CINECA through our expression of interest.






\bibliographystyle{mnras}
\bibliography{bibbcg}


%
%
%


\bsp	
\label{lastpage}
\end{document}